\documentclass[11pt,reqno]{amsart}

\newcommand{\aaa}{{a}}

\usepackage{amscd,amssymb,amsmath,amsthm}
\usepackage[arrow,matrix]{xy}
\usepackage{graphicx}
\usepackage{color}
\usepackage{cite}
\newtheorem{thm}{Theorem}

\newtheorem{pro}{Proposition}
\newtheorem{rk}{Remark}

\numberwithin{equation}{section} 

\newcommand{\bea}{\begin{eqnarray}}
\newcommand{\eea}{\end{eqnarray}}

\def\b{\beta}

\begin{document}
\title[New Phase transitions of the Ising model]{New
Phase transitions of the Ising model on Cayley trees}

\author{D. Gandolfo, F.H. Haydarov,  U. A. Rozikov, J. Ruiz}

 \address{D.\ Gandolfo and J.Ruiz\\Centre de Physique Th\'eorique, UMR 6207,Universit\'es Aix-Marseille
 et Sud Toulon-Var, Luminy Case 907, 13288 Marseille, France.}
\email {gandolfo@cpt.univ-mrs.fr\ \ ruiz@cpt.univ-mrs.fr}

\address{F.\ H.\ Haydarov\\ National University of Uzbekistan,
Tashkent, Uzbekistan.}
\email {haydarov\_imc@mail.ru}

 \address{U.\ A.\ Rozikov\\ Institute of mathematics,
29, Do'rmon Yo'li str., 100125, Tashkent, Uzbekistan.}
\email {rozikovu@yandex.ru}

\begin{abstract}

We show that  the nearest neighbors Ising  model on the Cayley tree exhibits new temperature driven phase transitions.
These transitions holds at various inverse temperatures different from the critical one.
They are depicted by a change in the number of Gibbs states  as well as by a drastic change of the behavior of free energies at these new transition points.
 \newline
 We also consider the model in presence of an external field and  compute the  free energies of translation invariant  periodic boundary conditions.
\end{abstract}
\maketitle



\section{Introduction}

Besides of its simplicity, the well known nearest neighbors (n.n.) Ising model on the Cayley tree  still offers new interesting phenomenon (see e.g.
 \cite{R} for a recent review).


On this tree,  as a consequence of non-amenability, not only Gibbs measures but also free energies depend on the boundary conditions.

A study of this dependence is given in \cite{GRRR}.
 It is shown there that for all previously known boundary conditions the free energies exist.


 When trying to complete this  study we discover new phase transition phenomenon that are the subject of the present paper.

Let $\Gamma^k= (V , L)$ be the uniform Cayley tree, where each vertex has $k + 1$ neighbors with $V$ being the set of vertices and $L$ the set of edges.

 The n.n. Ising model  is then  defined by the
 formal Hamiltonian
\begin{equation}
\label{h}
H(\sigma)=-J\sum_{\langle x,y\rangle\subset V}\sigma(x)\sigma(y)-B\sum_{x\in V}\sigma(x).
\end{equation}
Here the first sum runs over   n.n. vertices
$\langle x,y\rangle$, the spins $\sigma(x)$ take values $\pm 1$, and the real parameters  $J$ and  $B$ stand respectively for  the interaction energy and the magnetic field.

On this tree, there is a natural distance to be denoted by $d(x,y)$,
 being
 the number of nearest neighbor pairs  of the minimal path between  the vertices $x$ and $y$
 (by  path one means   a collection of  nearest neighbor pairs, two consecutive pairs
 sharing at least a given vertex).

For a fixed $x^0\in V$, the root,
let
\begin{equation*}
W_n=\{x\in V :\, d(x,x^0)=n\}, \ \
V_n=\{x\in V :\, d(x,x^0)\leq n\}
\end{equation*}
be respectively the sphere and the ball
of radius $n$ with center at $x^0$,  and for
$ x\in W_n$ let
$$
S(x)=\{y\in W_{n+1} :  d(y,x)=1\}
$$
be the set of direct successors of $x$.

 The (finite-dimensional) Gibbs distributions over configurations    at
 inverse temperature   $\beta=1/T$
 are defined by
 \begin{equation}\label{*}
\mu_n(\sigma_n)=Z^{-1}_n(h)
\exp\Big\{\beta J\sum_{\langle x,y\rangle\subset V_n} \sigma(x)\sigma(y)+\beta B\sum_{x\in V_n}\sigma(x)
+\sum_{x\in W_n}h_x\sigma(x)\Big\}
\end{equation}
  with partition functions given by
 \begin{equation}\label{pf}
Z_n(h)
=
\sum_{\sigma_n }
\exp\Big\{\beta J
\sum_{\langle x,y\rangle\subset V_n} \sigma(x)\sigma(y)+\beta B\sum_{x\in V_n}\sigma(x)
+\sum_{x\in W_n}h_x\sigma(x)\Big\}.
\end{equation}
Here the spin configurations $ \sigma_n $ belong to $\{-1,+1\}^{V_n}$
and
\begin{equation}
h=\{h_x\in R, \ \ x\in V\}
\end{equation}
is a collection of real numbers that stands for (generalized) boundary condition.

The probability distributions (\ref{*}) are said compatible if for all
$\sigma_{n-1}$
\begin{equation}\label{**}
\sum_{\omega_n}\mu_n(\sigma_{n-1}, \omega_n)=\mu_{n-1}(\sigma_{n-1})
\end{equation}
where the configurations
$\omega_n$ belong to $\{-1,+1\}^{W_n}$.

It is well known that this compatibility condition is satisfied if and only if for any $x\in V$
the following equation holds
\begin{equation}\label{***}
 h_x=\sum_{y\in S(x)}f_{\theta}(h_y+\beta B),
\end{equation}
where
\begin{equation}
\label{****}
\theta=\tanh(\b J), \quad f_{\theta}(h)
={\rm arctanh}(\theta\tanh h).
\end{equation}

Namely,   for any boundary condition
satisfying the functional equation (\ref{***}) there exists a unique Gibbs measure, the correspondence being one-to-one.

A boundary condition satisfying (\ref{***}) is called \emph{compatible}

The paper is organized as follows.  The results are given in Section~2.
Subsection 2.1 presents  the construction  of  new Gibbs measures and the computation of their corresponding free energies.
In Subsection 2.2 we examine the  behavior of free energies  of translation invariant and periodic b.c. with respect to an external field.
Proofs are given in Section~3.

\section{Results}

\subsection{Alternating Gibbs measures}

Here we restrict to a vanishing external field ($B=0$) and consider the half tree.
Namely the root $x^0$
has $k$ nearest neighbors.

We construct below new solutions of the functional equation (\ref{***}).
Namely, we let $q$ and $r$  be non-negative integers such that
$1\leq q \leq k-1$, $0\leq r\leq k$ and $r$ has the same parity as $k$.

Consider the  boundary condition
$h=\{h_x, x\in V\}$ with fields taking values $0,\pm h_1, \pm h_2$ defined by the following steps:

\begin{itemize}
\item[(i)]
if at vertex $x$ we have $h_x=0$, then the function has values
\begin{equation*}
\left\{\begin{array}{cl}
0 &  {\rm on}    \quad  r  \quad  {\rm vertices  \ of }
\quad S(x)
\\[3mm]
h_1 & {\rm on \ half \ of \ remaining \ vertices}
\\[3mm]
-h_1 & {\rm on  \ the \ remaining \ vertices};
\end{array}
\right.
\end{equation*}
\item[(ii)]
if at vertex $x$ we have $h_x=h_1$ (resp. $-h_1$) then on $q$ vertices  in $S(x)$ the function has value $h_2$ (resp. $-h_2$) and on other vertices, it takes value $0$;
\item[(iii)] if at vertex $x$ we have $h_x=h_2$ (resp. $-h_2$) then on each vertices in $S(x)$ the function has value $h_1$ (resp. $-h_1$).
\end{itemize}

It is easy to see that the boundary conditions in the above construction are compatible iff $h_1$ and $h_2$ satisfy the following system of equations:
\begin{equation}\label{h12}
\left\{\begin{array}{ll}
h_1=qf_{\theta}(h_2)\\[3mm]
h_2=kf_{\theta}(h_1)
\end{array}
\right.
\end{equation}
where $1\leq q\leq k-1$.

Denote
\begin{equation}
\theta_{\rm c}=1/\sqrt{qk}
\end{equation}

\begin{thm}\label{t1}
The system of equations (\ref{h12}) has a unique solution $(0,0)$, if $-\theta_{\rm c}\leq \theta\leq \theta_{\rm c}$ and three distinct solutions $(0,0)$,  $( h^*_1,  h^*_2)$, and $( -h^*_1,  -h^*_2)$  ($h^*_1,\, h^*_2 >0$), when $|\theta| >\theta_{\rm c}$.
\end{thm}

As a consequence of Theorem \ref{t1}, we can construct new Gibbs measures, say $\nu^{\rm ALT}_{h_1, h_2}$, with these solutions.
These measures also depend on the choice of the value of the root, and obviously
 differ for $|\theta| >\theta_{\rm c}$.

This phase transition can also be observed throughout the behaviour of the free energies.

Let us recall that the free energy of a compatible boundary condition (b.c.) is  defined as the limit:
\begin{equation}\label{fe1}
F(h)=-\lim_{n\to \infty}\frac{1}{ \beta |V_n|}
\ln Z_n( h)
\end{equation}
if it exists.
Hereafter $|\cdot|$ denotes  the cardinality of a set.

\begin{pro}\label{tfe}
\begin{itemize}
\item[a)] If $r\ne 0$ then free energies $F_{\rm ALT}$ of b.c.
(\ref{h12}) equal to
\begin{equation}
\label{odd}
\frac{k(k-q)}{(2k^2-rk-rq)}
\left(\aaa(0)
+
\frac{k-r}{ k-q}
\aaa(h_1)
+
\frac{q(k-r)}{ k(k-q)}
\aaa(h_2)\right),
\end{equation}
where
\begin{equation}
\label{at}
\aaa (t) =
-\frac{1}{2\beta}
\ln[4
\cosh(t +\b J)\cosh(t -\b J)].
\end{equation}

\item[b)] If $r=0$ then free energies has the following accumulating points
\begin{equation}
\label{F0}
\begin{cases}
\displaystyle
\frac{k-q}{ k+1}
\aaa(0)
+\frac{1}{ k+1}\aaa(h_1)
+
\frac{q}{ k+1}\aaa(h_2),
&\text{if}
\quad n=2m, \, m\to\infty\\[3mm]
\displaystyle
 \frac{k-q}{k(k+1)}
 \aaa(0)+
 \frac{k}{k+1}
 \aaa(h_1)
 +
 \frac{q}{ k(k+1)}\aaa(h_2),
&\text{if}
\quad n=2m+1
 \end{cases}
 \end{equation}
when the root takes value $0$, $\pm h_2$
and
\begin{equation}\label{F1}
\begin{cases}
\displaystyle
\frac{k-q}{ k(k+1)}\aaa(0)+
\frac{k}{ k+1}\aaa(h_1)+
\frac{q}{ k(k+1)}\aaa(h_2),
&
\text{if}
\quad
 n=2m,
 \\
 \displaystyle
\frac{k-q}{ k+1}\aaa(0)+\frac{1}{ k+1}\aaa(h_1)+
\frac{q}{ k+1}
\aaa(h_2),
&
\text{if}  \quad  n=2m+1
\end{cases}
 \end{equation} when the root takes value $\pm h_1$.
\end{itemize}
\end{pro}

 As it can be seen in Fig.~1, the free energies exhibit a drastic change at $\beta_c = {\rm arctanh}\, \theta_c$.
 This phase transition occurs at an inverse temperature greater than the critical temperature
$\beta_{\rm cr} = {\rm arctanh}\, 1/ k$.
Let us recall that above this inverse temperature, the equation (\ref{***}) has three different translation invariant solutions,
$\pm h^*,0$, where $h^*$ is the solution of $h=kf_{\theta}(h)$.
\begin{center}
\includegraphics[width=10cm]{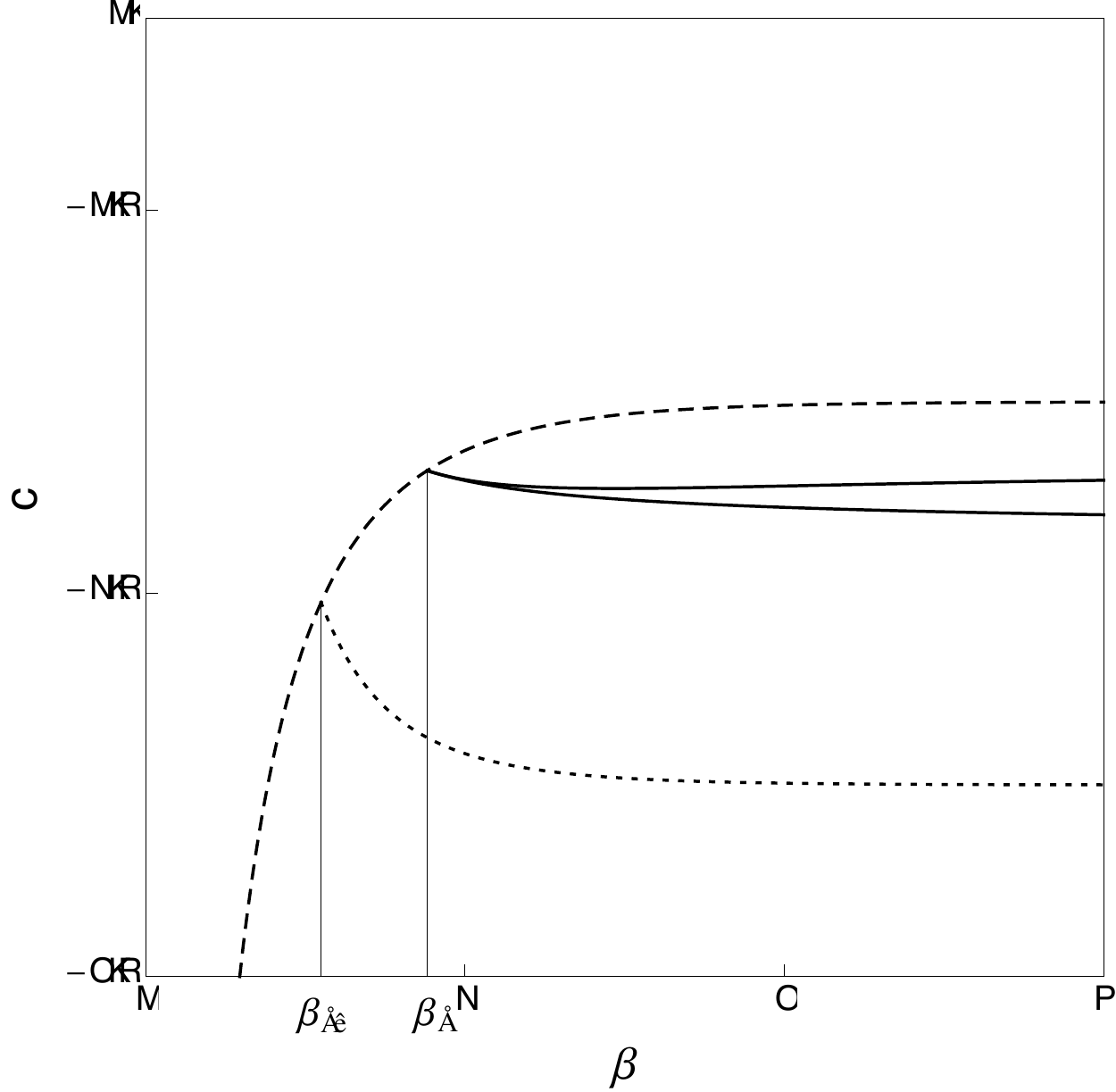}

{\footnotesize
\noindent
Fig.~1.
The free energies  of alternating b.c.: the two accumulating points
(\ref{F0}) (solid lines) for solutions
$( h^*_1,  h^*_2)$ (or $( -h^*_1,  -h^*_2)$); the dashed line represents the  free energy of the solution $(0,0)$.
Dotted line gives the free energy $F_{\rm TI}(h^*)$ of the translation invariant b.c. $h^*$.}
\end{center}

The above picture corresponds to the case $k=2$, $q=1$ for which
 one can easily solve the system (\ref{h12})
and get the following explicit solution:
\begin{equation}
\label{solh12}
\begin{cases}
\displaystyle
h^*_1
=
\frac{1}{ 2}
\ln
\left(
\frac{1+\theta}{ 1-\theta}
\cdot
\frac{\theta\sqrt{2\theta^2-1}+\theta^2+\theta-1
}{ \theta\sqrt{2\theta^2-1}-\theta^2+\theta+1}
\right)
\\[6mm]
\displaystyle
h^*_2
=
\frac{1}{ 2}
\ln
\frac{2\theta^2\sqrt{2\theta^2-1}+\theta^4+2\theta^2-1
}{
(1-\theta^2)^2}
\end{cases}
\end{equation}
and
$h_*=\ln
\frac{\theta+\sqrt{2\theta-1}
}{
1-\theta}
$;  $J=1$.


\begin{rk}
 Let us mention  the absence of residual entropy for alternating b.c. The residual entropy at $T=0$ is defined as the limit
\begin{equation}
S_{\infty}=-\lim_{\beta\to\infty}
\frac{F(h)-F_{\infty}
}{
(1/\beta)},
\end{equation}
where $F_{\infty}=\lim_{\beta\to\infty}F(h)$,
and it can be easily seen from the formulae of Proposition~\ref{tfe} that it vanishes for such b.c..

Namely, the corresponding Gibbs measures tends in the zero temperature limit to a precise ground state configuration.
This ground state configuration is a weakly periodic one in the sense of ref.~\cite{GRS}. This means that  the spins of endpoints of a subset  $D \subset L$ alternate sign although they did not otherwise.
\end{rk}

\begin{rk}
The construction above can be easily generalized.
Namely, one can let the parameters $q$ and $r$ vary depending on the generation $n$ or even more on the
position of the vertex.
\end{rk}

\begin{rk}
Although, it was proved in \cite{GRRR} that in absence of external field, the free energy exists for known b.c.,  the statement b) of Proposition~\ref{tfe} shows that
 of free energy of alternating Gibbs measures may not exist.

The nonexistence  of free energy in the sense that it  may assume different accumulating points
 already appears when a magnetic field is turned on. This is discussed in the next subsection.
\end{rk}

\subsection{The influence of a magnetic field}

Here we consider the model in presence of an external field ($B\ne 0$) on the complete tree and discuss
the behavior of free energies as function of that field.

Let us first consider
translation invariant boundary conditions.

\begin{pro}\label{fB}

The free energies of  compatible translation-invariant  b.c. exist and are given by
\begin{equation}\label{fB1}
F_{\rm TI}( h)
=-\frac{1}{2\beta}\ln[4\cosh(h+\b(B+J))
\cosh(h+\b(B-J))],
\end{equation}
where $h$ are the solutions of
the equation
\begin{equation}\label{ti}
h=kf_{\theta}(h + \beta B).
\end{equation}
\end{pro}

Let us recall that in the ferromagnetic case ($J>0$), there exits
$B^F(\beta)$ given by the equation
 \begin{equation}
 \label{BFerro}
 B^F(\beta)=
 (1/\beta)
 \left[k
 \operatorname{arctanh}
\left(
\frac{k \theta -1}{k/ \theta -1}
\right)^{1/2}
-
 \operatorname{arctanh}
\left(
\frac{k-1/\theta}{k-\theta}
\right)^{1/2}
 \right]
 \end{equation}
such that the equation (\ref{ti}) has
\begin{enumerate}
\item[(i)]
a unique solution for
$\theta \leq \theta_{\rm cr}
=1/k
$
or for $\theta > \theta_{\rm cr}$ and
$|B|>B^F(\beta)$,
\item[(ii)]
two distinct solutions for
$\theta >\theta_{\rm cr}$ and
$|B|=B^F(\beta)$,
\item[(iii)]
three distinct solutions, say
$h_{\operatorname{min}}$,
$h_0$
and
$h_{\operatorname{max}}$,
if
$\theta > \theta_c$ and
$|B|<B^F(\beta)$.
\end{enumerate}
See   \cite{Ge} chapter 12, or e.g \cite{BRZ}.

Some particular plots of these free energies as functions of the magnetic field are shown in Fig.~2.

\begin{center}
\includegraphics[width=10cm]{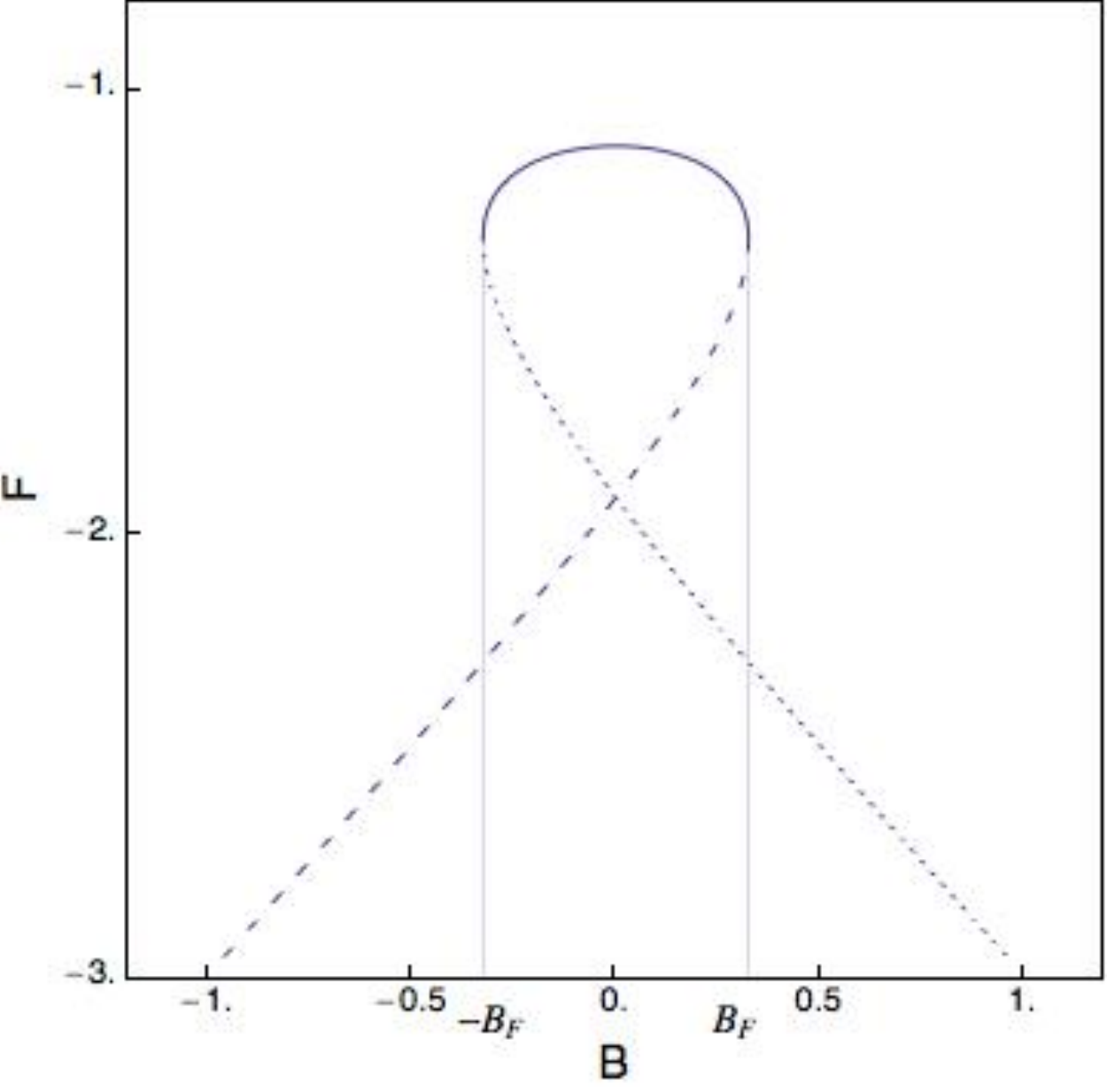}

{\footnotesize \noindent
 Fig.~2.
 The free energies of T.I. b.c. as functions of the magnetic field: $F_{\rm TI}(h_{\operatorname{min}})$ (dashed lines), $F_{\rm TI}(h_0)$
 (solid lines), and $F_{\rm TI}(h_{\operatorname{min}})$ (dotted lines).}
\end{center}

In order to draw these  free energies
as
functions of the magnetic field $B$, we  notice that
the equation (\ref{ti})  gives
\begin{equation}\label{Bdeh}
B(h) =-\frac{h}{\beta}+
\frac{
1
}{
2\beta}
\ln \frac{\sinh(J\beta+h/k)}{\sinh(J\beta-h/k)}
\end{equation}
We then use the  parametric representation defined by the following mapping:
\begin{equation*}
  h \to (B(h), F(h)),
\end{equation*}
where the function $F(h)$ is defined by inserting
(\ref{Bdeh}) into (\ref{fB1}).

A vertex on Cayley tree is called even (resp. odd) when its  distance to the root is even (resp. odd).

We  now turn to
periodic boundary conditions.
A periodic b.c.  $\{h, h'\}$ defined  by
\begin{equation}
h_x=
\left\{\begin{array}{ll}
h\ \ \mbox{if $x$ is  even}\\[3mm]
h'\ \ \mbox{if} \ \  \mbox{$x$ is odd}
\end{array}\right.
\end{equation}
is compatible iff
\begin{equation}\label{per}
h=kf_{\theta}(h'+ \b B), \ \
h'=k f_{\theta}(h+ \b B).
\end{equation}

\begin{pro}\label{fBP}
The free energies of periodic b.c. are given by
\begin{equation}\label{fB2}
F_{\rm Per}(\{h, h'\})=
-\frac{1}{ k+1}
\left\{\begin{array}{ll}
k d(h)+ d(h'),\ \ \mbox{if} \ \ n=2m, \ \ m\to\infty\\[3mm]
d(h)+k d(h'), \ \ \mbox{if} \ \ n=2m+1,
\end{array}\right.
\end{equation}
where
\begin{equation}\label{dt}
d(t)=
\frac{1}{2\beta }
\ln\Big( 4\cosh[t+\b(B+J)]\cosh[t+\b(B-J)]\Big).
\end{equation}
\end{pro}

In the antiferromagnetic case ($J<0$),
there exits
$B^{\rm AF}(\beta)$ given by the equation
 \begin{equation}
 \label{AFerro}
 B^{ AF}(\beta)=
 (1/\beta)
 \left[
 k
  \operatorname{ arctanh}
\left(
\frac{k \theta -1}{k/ \theta -1}
\right)^{1/2}
+
 \operatorname{ arctanh}
\left(
\frac{k -1/\theta}{k -\theta}
\right)^{1/2}
 \right]
 \end{equation}
  such that
  the system of equations (\ref{per}) has

  \begin{enumerate}
\item[(i)]
a unique solution for
$
\theta \geq -\theta_c
$
or for $\beta < -\theta_c$ and
$|B|>B^{AF}(\beta)$,
\item[(ii)]
two distinct solutions
 say
$\{\bar{h},\bar{h}'\}$
and
$\{\bar{h'},\bar{h}\}$
($\bar{h}<\bar{h}'$)
if
$\theta < -\theta_c$ and
$|B|<B^{AF}(\beta)$.
\end{enumerate}
See e.g.  \cite{Ge} chapter 12 or \cite{BRZ}.
It is clear from  (\ref{fB2})
that the first accumulating point for
$\{\bar{h},\bar{h}'\}$
equals  the second one of
$\{\bar{h'},\bar{h}\}$, and vice-versa.

\begin{rk}
It is likely that the alternating Gibbs measures will survive in presence of a small magnetic field.
\end{rk}


%
%

\section{Proofs}

\subsection{Proof of Theorem \ref{t1}}

From system (\ref{h12}) we get $u=g(u)$ with $u=h_2$ and $g(u)=kf_\theta(qf_\theta(u))$. It is easy to see that
$g(0)=0$, $g(-u)=-g(u)$, $g'(0)=kq\theta^2$ and $g(u)$ is a bounded, increasing function of $u$. It is a concave function for $u>0$. Under $|g'(0)|>1$ there is a sufficiently small neighborhood of $u=0$: $(-\varepsilon, +\varepsilon)$ such that $g(u)<u$, for $u\in (-\varepsilon, 0)$ and $g(u)>u$, for $u\in (0, +\varepsilon)$. By above mentioned properties of $g(u)$, there is a unique solution $u^*>0$ between 0 and $+\infty$, by oddness of $g$, the number $-u^*$ is a unique solution between $-\infty$ and $0$. If $|g'(0)|\leq 1$ then $g(u)<u$ for $u>0$ and $g(u)>u$ for $u<0$, so $u=0$ is the unique solution in this case.

%
%
%

 \subsection{Proof of Proposition \ref{tfe}}

For compatible  boundary conditions,
the free energy is given by the
formula
\begin{equation}\label{fe2}
F( h)=\lim_{n\to\infty}\frac{1}{ |V_n|}
\sum_{x\in V_{n}}a(h_x)
\end{equation}
with the function $a$ defined by (\ref{at}),
see \cite{GRRR}.


%

 Put
$$\alpha_n=|\{x\in W_n: h_x=0\}|;$$
\begin{equation}\label{ss}
\beta_n=|\{x\in W_n: h_x=h_1\}|; \ \ \gamma_n=|\{x\in W_n: h_x=-h_1\}|;
\end{equation}
$$\delta_n=|\{x\in W_n: h_x=h_2\}|; \ \ \xi_n=|\{x\in W_n: h_x=-h_2\}|.$$
(Recall that $W_n$ is the sphere with the center $x^0$ and radius $n$ on the half tree).

 \subsubsection{{\rm a)} $r\ne 0$.}

In this case the following recurrence system holds
\begin{equation}\label{oab}
\left\{\begin{array}{lllll}
\alpha_{n+1}=r\alpha_n+(k-q)(\beta_n+\gamma_n)\\[3mm]
\beta_{n+1}=
\frac{k-r
}{
2}
\alpha_n+k\delta_n\\[3mm]
\gamma_{n+1}=
\frac{k-r
}{
2}
\alpha_n+k\xi_n\\[3mm]
\delta_{n+1}=q\beta_n,\\[3mm]
\xi_{n+1}=q\gamma_n.
\end{array}\right.
\end{equation}

Denoting $w_n=\beta_n+\gamma_n$ and $v_n=\delta_n+\xi_n$ from (\ref{oab})
we get
\begin{equation}\label{oa1}
\left\{\begin{array}{lll}
\alpha_{n+1}=r\alpha_n+(k-q)w_n\\[3mm]
w_{n+1}=(k-r)\alpha_n+kv_n\\[3mm]
v_{n+1}=qw_n.
\end{array}\right.
\end{equation}
From the first equation of (\ref{oa1}) we get
$$
w_n=
\frac{\alpha_{n+1}-r\alpha_n
}{
k-q}
$$
which by second and third equations gives
\begin{equation}\label{oa2}
\alpha_{n+2}-r\alpha_{n+1}-(k^2-rk+rq)\alpha_n+rkq\alpha_{n-1}=0.
\end{equation}

The characteristic equation for (\ref{oa2}) has the following form (setting $\alpha_n=\lambda^n $):
$$
\lambda^{3}-r\lambda^{2}-(k^2-rk+rq)\lambda+rkq=0,
$$
which has solutions
$$\lambda_1=k, \ \  \lambda_{2,3}=
\frac{-(k-r)\pm\sqrt{(k-r)^2+4rq}
}{
 2}
$$
Then the general solution to (\ref{oa2}) is
\begin{equation}\label{og}
\alpha_n=c_1\lambda_1^n+c_2 \lambda_2^n
+c_3\lambda_3^n,
\end{equation}
where the coefficients $c_1$, $c_2$, $c_3$ are determined
by the initial conditions on $\alpha_n$, $n=0,1,2$.

By taking into account the system (\ref{oa1}) we get
$$w_n=
\frac{1
}{
k-q}
\left(c_1(\lambda_1-r)\lambda_1^n+c_2(\lambda_2-r)\lambda_2^n+c_3(\lambda_3-r)\lambda_3^n\right)
$$
and
$$v_n
=
\frac{q
}{
k-q}
\left(c_1(\lambda_1-r)\lambda_1^{n-1}
+c_2(\lambda_2-r)\lambda_2^{n-1}
+c_3(\lambda_3-r)\lambda_3^{n-1}
\right).
$$
Denote
\begin{equation}\label{ABC}
A_n=\sum_{i=0}^n\alpha_i, \ \ B_n=\sum_{i=0}^n w_i, \ \ C_n=\sum_{i=0}^n v_i.
\end{equation}
For $r\ne 0$ it is easy to check that $|\lambda_{2,3}|<k$.
Moreover, for half tree we have
$$|V_n|=
\frac{k^{n+1}-1
}{
k-1}.$$
By these inequalities, in case $r\ne 0$, for (\ref{ABC}) we obtain
$$\lim_{n\to\infty}
\frac{A_n
}{
|V_n|}
=c_1, \ \ \lim_{n\to\infty}
\frac{B_n
}{
|V_n|}
=
\frac{c_1(k-r)
}{ k-q},
 \ \
\lim_{n\to\infty}
\frac{C_n
}{
|V_n|}
=
\frac{c_1q(k-r)
}{
k(k-q)}.$$

Hence the  free energy of alternating b.c. are given by:
\begin{equation}\label{F0o}
F_{\rm ALT}=
\lim_{n\to \infty}
\frac{1}{|V_n|}
\left(A_n\aaa(0)+B_n\aaa(h_1)+C_n\aaa(h_2)\right)=
\end{equation}
$$c_1
\left(\aaa(0)+
\frac{k-r}{ k-q}
\aaa(h_1)+
\frac{q(k-r)
}{ k(k-q)}
\aaa(h_2)\right).$$

Hence the limit exists, moreover its an explicit value is obtained.

Now let us  compute $c_1$ in the different cases:

{\it The root take value $0$:} In this case $\alpha_0=1$, $\alpha_1=r$ and $\alpha_2=r^2+(k-r)(k-q)$. Consequently, for
$c_1, c_2$ and $c_3$ we have

\begin{equation}\label{cccc}
\left\{\begin{array}{lll}
c_1+c_2+c_3=1\\[3mm]
c_1\lambda_1+c_2\lambda_2+c_3\lambda_3=r\\[3mm]
c_1\lambda_1^2+c_2\lambda_2^2+c_3\lambda_3^2=r^2+(k-r)(k-q).
\end{array}\right.
\end{equation}
From this system we get
\begin{equation}\label{c1}
c_1=
\frac{k^2-kq
}{
2k^2-rk-rq}.
\end{equation}

{\it The root take value $\pm h_1$:} In this case $\alpha_0=0$, $\alpha_1=k-q$ and $\alpha_2=r(k-q)$. Consequently, for
$c_1, c_2$ and $c_3$ we have

\begin{equation}\label{cccd}
\left\{\begin{array}{lll}
c_1+c_2+c_3=0\\[3mm]
c_1\lambda_1+c_2\lambda_2+c_3\lambda_3=k-q\\[3mm]
c_1\lambda_1^2+c_2\lambda_2^2+c_3\lambda_3^2=r(k-q).
\end{array}\right.
\end{equation}
It is surprise that from this system we again get $c_1$ given by (\ref{c1}).

{\it  The root take value $\pm h_2$:} In this case $\alpha_0=0$, $\alpha_1=0$ and $\alpha_2=k(k-q)$. Consequently, for
$c_1, c_2$ and $c_3$ we have

\begin{equation}\label{equainc1}
\left\{\begin{array}{lll}
c_1+c_2+c_3=0\\[3mm]
c_1\lambda_1+c_2\lambda_2+c_3\lambda_3=0\\[3mm]
c_1\lambda_1^2+c_2\lambda_2^2+c_3\lambda_3^2=k(k-q).
\end{array}\right.
\end{equation}
Again it is surprise that from this system we get $c_1$ given by (\ref{c1}).

Hence for $r\ne 0$, substituting $c_1$ given by (\ref{c1}) in (\ref{F0o}) we obtain (\ref{odd}).

 \subsubsection{{\rm b)} $r=0$.}

In this case $k$ is even. We shall show that the free energy exhibits different accumulating (limiting) points.

 We have the recurrence system 
\begin{equation}\label{gab}
\left\{\begin{array}{lllll}
\alpha_{n+1}=(k-q)(\beta_n+\gamma_n)\\[3mm]
\beta_{n+1}=
\frac{k
}{
2}\alpha_n+k\delta_n\\[3mm]
\gamma_{n+1}=
\frac{k
}{
2}
\alpha_n+k\xi_n\\[3mm]
\delta_{n+1}=q\beta_n\\[3mm]
\xi_{n+1}=q\gamma_n,
\end{array}\right.
\end{equation}
where $\alpha_0, \beta_0, \gamma_0, \delta_0, \xi_0\in \{0,1\}$ and $\alpha_0+\beta_0+\gamma_0+\delta_0+\xi_0=1$, i.e., only one of them equal 1 depending from which value $0$, $\pm h_1$ and $\pm h_2$ at the root we start our construction.

Since a free energy is an even function, we need only know $\alpha_n$,
$w_n=\beta_n+\gamma_n$ and $v_n=\delta_n+\xi_n$. By (\ref{gab}) we get $v_{n+1}=q w_n$. Thus the system (\ref{gab}) can be reduces to the following
\begin{equation}\label{ga}
\left\{\begin{array}{ll}
\alpha_{n+1}=(k-q)w_n\\[3mm]
w_{n+1}=k\alpha_n+kq w_{n-1},
\end{array}\right.
\end{equation}
with $\alpha_0+w_0=1$.

By (\ref{ga}) we have
$$w_{n+1}=k(k-q)w_{n-1}+kqw_{n-1}=k^2w_{n-1}, n=1,2,\dots$$
This gives
\begin{equation}\label{w}
w_{2m}=k^{2m}w_0, \ \ w_{2m+1}=k^{2m}w_1.
\end{equation}
Now using these equalities we obtain
\begin{equation}\label{a}
\alpha_{2m}=(k-q)k^{2m-2}w_1, \ \ \alpha_{2m+1}=(k-q)k^{2m}w_0.
\end{equation}
\begin{equation}\label{v}
v_{2m}=q k^{2m-2}w_1, \ \ v_{2m+1}=q k^{2m}w_0.
\end{equation}

Using (\ref{w})-(\ref{v}), for (\ref{ABC})  we get
 \begin{equation}\label{A}
A_{2m}=\alpha_0+
\frac{k-q
}{ k^2-1}
(w_0+w_1)(k^{2m}-1), \ \ A_{2m+1}=A_{2m}+(k-q)k^{2m}w_0.
\end{equation}
 \begin{equation}\label{B}
B_{2m}=
\frac{1
}{
k^2-1}\left(w_0(k^{2m+2}-1)+w_1(k^{2m}-1)\right), \ \
B_{2m+1}=B_{2m}+k^{2m}w_1.
\end{equation}
 \begin{equation}\label{C}
C_{2m}=v_0+
\frac{q
}{ k^2-1}
(w_0+w_1)(k^{2m}-1), \ \
C_{2m+1}=C_{2m}+qk^{2m}w_0.
\end{equation}

Using these equalities we get the corresponding free energy
\begin{equation}\label{gF}
F_{\rm ALT}=
\lim_{n\to \infty}\frac{1}{|V_n|}\left(A_n\aaa(0)+B_n\aaa(h_1)+C_n\aaa(h_2)\right).
\end{equation}

Now by (\ref{A})-(\ref{C}) from (\ref{gF}) we get
for $
F_{\rm ALT}$
the values
\begin{equation}\label{gF1}
\left\{\begin{array}{ll}
\frac{(k-q)(w_0+w_1)
}{ k(k+1)}
\aaa(0)
+
\frac{k^2w_0+w_1
}{ k(k+1)}
\aaa(h_1)
+
\frac{q(w_0+w_1)
}{ k(k+1)}
\aaa(h_2), \ \ \mbox{if} \ \ n=2m, \, m\to\infty\\[5mm]
 \frac{(k-q)(k^2w_0+w_1)
 }{ k^2(k+1)}
 \aaa(0)
 +
 \frac{w_0+w_1
 }{ k+1}
 \aaa(h_1)+
 \frac{(k^2w_0+w_1)
 }{ k^2(k+1)}
 \aaa(h_2), \ \ \mbox{if} \ \ n=2m+1.
 \end{array}\right.
 \end{equation}

In case  the root takes value $0$, we have $w_0=0, \, w_1=k$, which by (\ref{gF1}) gives
(\ref{F0}).

 In case  the root takes value $\pm h_1$, we  have $w_0=1, \, w_1=0$, which by (\ref{gF1}) gives
the corresponding free energy (\ref{F1}). Note that even though the two values do not coincide,  they exchange their values depending on the parity of $n$.

In case  the root takes value $\pm h_2$, we have $w_0=0, \, w_1=k$, and consequently we get (\ref{F0}).

This ends the proof of Proposition \ref{tfe}.

\subsection{Proof of Proposition \ref{fB}  and \ref{fBP}}

In presence of  a magnetic field
the free energies are given by the
formula
\begin{equation}\label{fe3}
F(h)=-\lim_{n\to\infty}\frac{1}{ |V_n|}
\sum_{x\in V_{n}}d(h_x)
\end{equation}
where
$d(t)$ is defined by (\ref{dt}): see \cite{GRRR} or (\ref{fe2}), and take into  account
the field $B$. 

This  formula provides
the relation (\ref{fB1}) for the free energies of translation-invariant
b.c.,
and leads for
 periodic b.c. to:
\begin{equation}\label{36}
F_{\rm Per}(\{h,h'\})
=-\lim_{n\to\infty}
\frac{1}{ |V_n|}
\left(
|V_{{\rm even},n}|d(h)+|V_{{\rm odd},n}|d(h')
\right).
\end{equation}
Here, $V_{{\rm even},n}$ (resp. $V_{{\rm odd},n}$) denote the even (resp. odd) sites of $V_n$.

  Then the following formulas
$$
|W_n|=(k+1)k^{n-1}, \ \
|V_n|=
\frac{(k+1)k^n-2
}{ k-1}
$$
\begin{equation}\label{VW}
|V_{{\rm even},2m}|=\sum_{i=0}^m |W_{2i}|=1+
\frac{k(k^{2m}-1)
}{ k-1},
\ \
|V_{{\rm odd},2m+1}|=\sum_{i=0}^m |W_{2i+1}|=
\frac{k^{2m+2}-1
}{k-1}
\end{equation}
imply (\ref{fB2}).

\section*{ Acknowledgements}

U.A. Rozikov thanks the Universit\'e du Sud Toulon Var for financial support and the Centre de Physique Th\'eorique--Marseille for kind hospitality.

\end{document}